\documentclass[11pt]{article}
\usepackage[utf8]{inputenc}

\usepackage{amsmath,amsfonts,graphicx}
\usepackage{fullpage}
\usepackage{lineno}
\usepackage{xcolor}
\usepackage{pdflscape}
\usepackage{hyperref}
\usepackage{multicol,multirow}
\usepackage{authblk}
\usepackage{rotating}
\usepackage[backend=biber, style=numeric]{biblatex}
\usepackage{tikz}
\usetikzlibrary{positioning, arrows.meta, decorations.pathreplacing, shapes.geometric, arrows.meta, positioning, calc, shadows.blur, backgrounds, fit}

\title{Constructing Contact and Connectivity Matrices for Infectious Disease Modelling}

\author[1,2]{Xiahui Li}
\author[3]{Dongni Zhang}
\author[4]{Neha Bansal}
\author[5]{Jessica R.E. Bridgen}
\author[5]{Chris Jewell} 
\author[6]{Emma McBryde}
\author[7]{Glenn Marion}
\author[8]{Emily Nixon}
\author[9]{Philip D. O'Neill}
\author[10]{David J. Pascall}
\author[11]{Lorenzo Pellis}
\author[12]{Simon E.F. Spencer}
\author[13]{Panayiota Touloupou}
\author[5]{Lloyd Chapman}
\author[1,2]{Ben Swallow}

\affil[1]{School of Mathematics and Statistics, University of St Andrews, St Andrews, UK}
\affil[2]{Centre for Research into Ecological and Environmental Modelling, University of St Andrews, St Andrews, UK }
\affil[3]{Department of Health, Medicine and Caring Sciences, Linköping University, Linköping, Sweden}
\affil[4]{School of Mathematics, Cardiff University, Cardiff, UK}
\affil[5]{School of Mathematical Sciences, Lancaster University, Lancaster, UK}
\affil[6]{Australian Institute of Tropical Health and Medicine, James Cook University, Townsville, Australia}
\affil[7]{Biomathematics and Statistics Scotland, Edinburgh, UK}
\affil[8]{Department of Mathematical Sciences, University of Liverpool, Liverpool, UK}
\affil[9]{School of Mathematical Sciences, University of Nottingham, Nottingham, UK}
\affil[10]{MRC Biostatistics Unit, University of Cambridge, Cambridge, UK}
\affil[11]{Department of Mathematics, University of Manchester, UK}
\affil[12]{Department of Statistics, University of Warwick, UK}
\affil[13]{School of Mathematics, University of Birmingham, Birmingham, UK}
\date{}

\begin{document}

\maketitle

*Corresponding author. Email: bts3@st-andrews.ac.uk and l.chapman4@lancaster.ac.uk

\begin{abstract}
Contact (or mixing, or more generally connectivity) matrices are a fundamental component of modelling and inference for infectious disease epidemiology. Their structure and parametrisation directly accounts for the frequency of interactions between different subpopulations of individuals, as well as having the potential to encode dynamic heterogeneity in these interactions across demographic axes, space and time. Considerable research has been devoted to the structure and estimation of (components of) these matrices to help inform outbreak control and forecast disease spread. In this paper, we review the existing literature on the data types used to construct contact matrices and the methods for incorporating uncertainties and heterogeneities into them. We also highlight remaining challenges and future directions in the use of these contact matrices for epidemiological research.
\end{abstract}

\section{Introduction}

The spread of infectious diseases through human populations is inherently heterogeneous, as individuals differ in their biology, behaviour, and social context, and these differences shape who infects whom \cite{mossong2008social, lloyd2005superspreading, liu2021rapid}. Capturing this heterogeneity is one of the main challenges of epidemic modelling, and connectivity matrices have been used to provide a more realistic representation of contact and movement patterns within and between populations \cite{prem2017projecting, mistry2021inferring}.

\begin{figure*}[h!]
    \centering
    \includegraphics[width=0.5\linewidth]{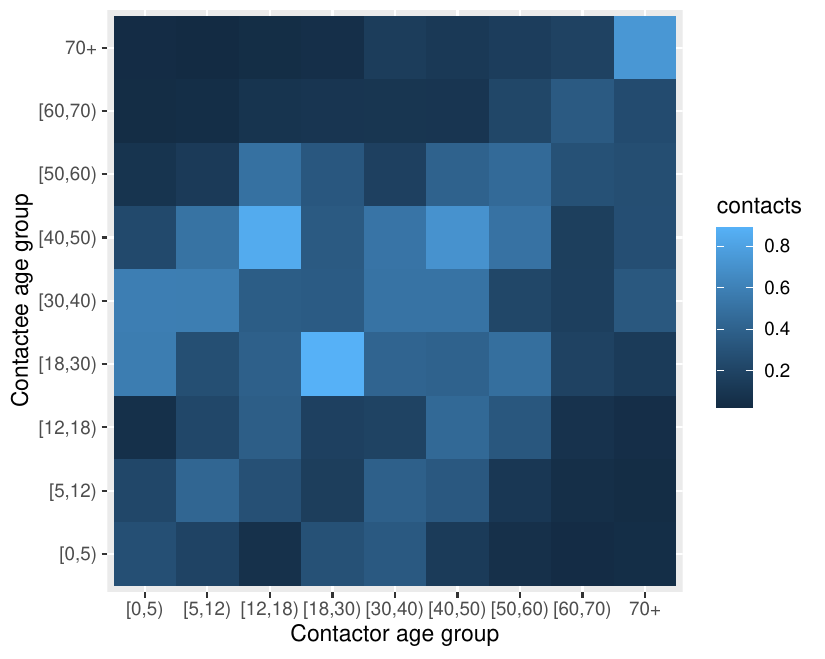}
    \caption{Example connectivity matrix. Colour shows the mean number of contacts each individual in a contactor age group (horizontal axis) has with individuals in different contactee age groups (vertical axis) per day in the first period of the CoMix contact survey (23rd March-3rd June 2020) conducted during the COVID-19 pandemic \cite{gimma2022comix}. Data obtained from \cite{gimma2024} and matrix calculated using the \texttt{contact\_matrix()} function from the \texttt{socialmixr} \texttt{R} package \cite{socialmixr}.}
    \label{fig:comix-contact-matrix}
\end{figure*}

A connectivity matrix is a mathematical tool used to quantify the average number of interactions between different subgroups of a population over a given period. For example, in a connectivity matrix $C = (c_{ij})$, each entry $c_{ij}$ represents the average number of contacts that each individual in group $i$ has with individuals in group $j$ per unit of time, where groups reflect any discrete stratification of the population (e.g. age groups, geographical locations, occupations, behavioural patterns, {\em etc.}) (Figure \ref{fig:comix-contact-matrix}). We distinguish this from a contact matrix, which typically refers to an empirically measured or survey-derived matrix for a specific stratification, most commonly age \cite{mossong2008social}. A connectivity matrix, by contrast, is a more general construct that allows any such discrete structure, including but not limited to age-structured contact matrices \cite{manna2024generalized}. For example, it may represent movement or mixing between geographic regions \cite{jewell2023bayesian}, or partnership changing rates for sexually transmitted infections \cite{walker2012revision}.

\begin{figure*}[h!]
    \centering
    \vspace{\baselineskip}
    \begin{tikzpicture}[
        state/.style={
            rectangle,
            draw,
            text centered,
            minimum height=1.3cm,
            minimum width=1.3cm,
        },
        arrow/.style={
            -Stealth,
            thick,
        }
    ]
    
        \node[state] (S) {$S_i$};
        \node[state, right=1.2cm of S] (I) {$I_i$};
        \node[state, right=1.2cm of I] (R) {$R_i$};
        \draw[arrow] (S) -- (I) node[midway, above] {$\lambda_i$};
        \draw[arrow] (I) -- (R) node[midway, above] {$\gamma$};
    \end{tikzpicture}
    \caption{Stratified Susceptible-Infectious-Recovered (SIR) model. $S_i,I_i,R_i$ denote the numbers of susceptible, infectious, and recovered individuals respectively in subgroup $i$ in the population (which may represent a particular age group or spatial region). $\lambda_i=p\sum_j c_{ij}\frac{I_j}{n_j}$ is the rate of infection for subgroup $i$, where $c_{ij}$ is the mean number of contacts each individual in subgroup $i$ has with individuals in subgroup $j$, $n_j$ is the number of individuals in subgroup $j$, $p$ is the probability of infection per contact, and $\gamma$ is the rate of recovery.}
    \label{fig:sir-model}
\end{figure*}
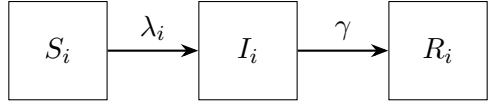

A simple example of how a connectivity matrix enables heterogeneity to be incorporated into a transmission model is the susceptible-infectious-recovered (SIR) model in Figure \ref{fig:sir-model}, in which the numbers of susceptible, infectious and recovered individuals in the population ($S$, $I$ and $R$) have been stratified into different subgroups denoted by subscript $i$. The force of infection acting on a single susceptible in subgroup $i$, $\lambda_i$, is given by:
$$
\lambda_i = p\sum_j c_{ij}\frac{I_j}{n_j},
$$
where the parameter, $p$, is the probability of infection per contact event (although it might be defined more generally simply as a scaling factor, as in certain contexts it might take values larger than 1, e.g.\ for partnership changing rates and SIS-type sexually-transmitted infections, for which repeated infections in the same partnership are possible) and $n_j$ is the population size of subgroup $j$.

\noindent This model can be described by the following system of ordinary differential equations: 
\begin{align*}
\frac{dS_i}{dt }&= -\lambda_i S_i\\
\frac{dI_i}{dt} &= \lambda_i S_i - \gamma I_i\\
\frac{dR_i}{dt} &= \gamma I_i,
\end{align*}
where $\gamma$ is the recovery rate. With this model, we are able to track the numbers of individuals in different infection states and in different subgroups of the population over time, and how these dynamics are affected by differences in contact rates between different subgroups (i.e.\ differences in the entries in $C=(c_{ij})$).

The construction of connectivity matrices is based on a diverse range of data sources. Social contact surveys have long served as the primary foundation \cite{mossong2008social, hoang2019systematic, coletti2020comix, goodfellow2025post}, while demographic and epidemiological data provide complementary information on the structure of the population and the mixing pattern \cite{diekmann2010construction, read2014social}, and spatial and mobility data capture geographic heterogeneity in contact behaviour \cite{oliver2020mobile,shubin2024influence, prestige2025estimating}. More recently, pathogen genomic data have emerged as a powerful means of inferring contact structure from the information of pathogen genome sequence data \cite{duplessis2021phylogeography, demaio2015structuredcoalescent, kuhnert2016mtbd}.

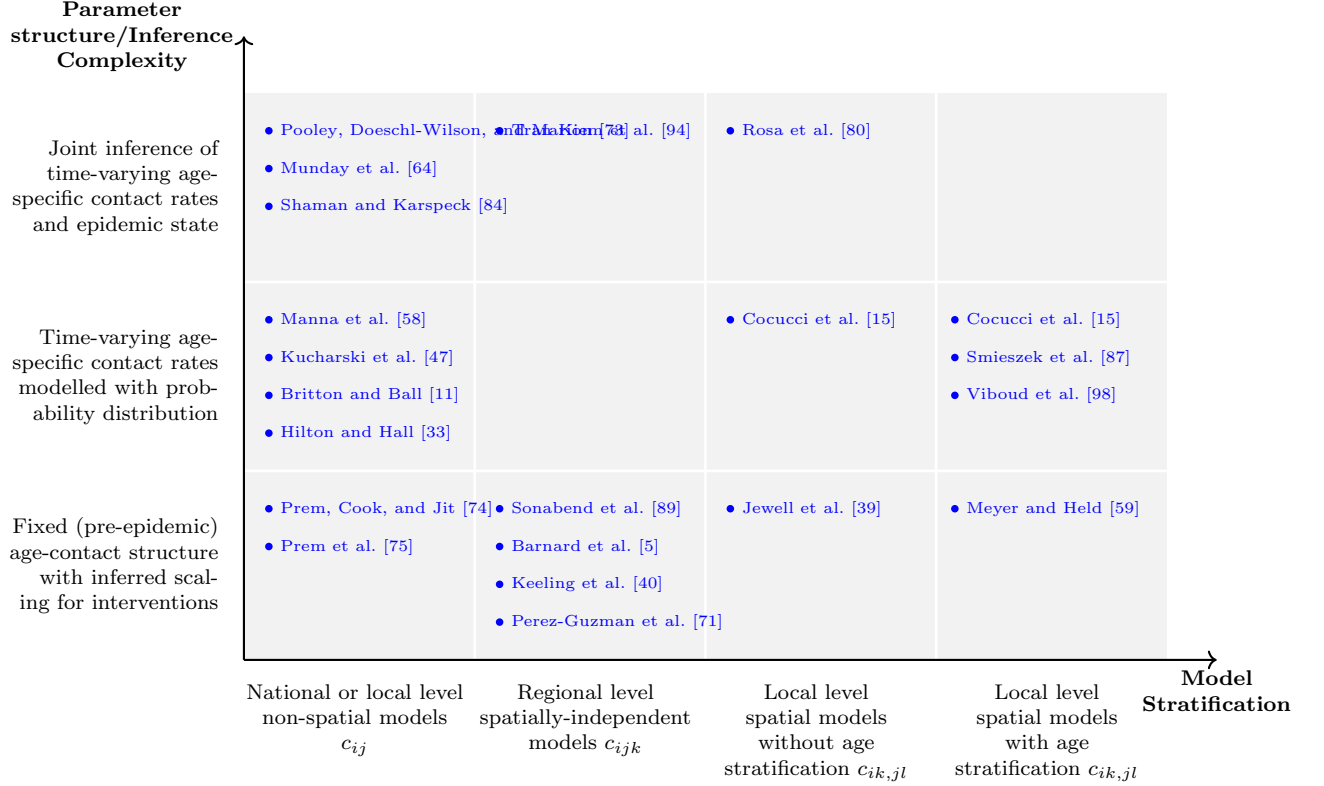
\begin{figure*}[h]
\centering
\begin{tikzpicture}[x=1.65cm, y=2.5cm]


\fill[gray!10] (0,0) rectangle (7.4,3);
\foreach \x in {1.85, 3.7, 5.55} \draw[white, line width=1pt] (\x,0) -- (\x,3);
\foreach \y in {1,2}   \draw[white, line width=1pt] (0,\y) -- (7.4,\y);
\draw[thick, ->] (0,0) -- (7.8,0) node[below, align=center, font=\bfseries\scriptsize] {Model\\Stratification};
\draw[thick, ->] (0,0) -- (0,3.3) node[left, align=center, font=\bfseries\scriptsize] {Parameter\\structure/Inference\\Complexity};
\tikzset{xlab/.style={
    below, 
    align=center, 
    font=\scriptsize, 
    text depth=0.5ex, 
    inner sep=5pt,
    yshift=-0.1cm
}}
\node[xlab] at (0.92, 0)  {
    \begin{tabular}{@{}c@{}} 
        National or local level \\ non-spatial models \\ $c_{ij}$ 
    \end{tabular}
};
\node[xlab] at (2.77, 0)  {
    \begin{tabular}{@{}c@{}} 
        Regional level \\ spatially-independent \\ models $c_{ijk}$ 
    \end{tabular}
};
\node[xlab] at (4.62, 0)  {
    \begin{tabular}{@{}c@{}} 
        Local level \\ spatial models \\ without age \\ stratification $c_{ik,jl}$ 
    \end{tabular}
};
\node[xlab] at (6.47, 0)  {
    \begin{tabular}{@{}c@{}} 
        Local level \\ spatial models \\ with age \\ stratification $c_{ik,jl}$ 
    \end{tabular}
};
\tikzset{ylab/.style={left, align=right, font=\scriptsize, text width=3.2cm, xshift=-0.2cm}}
\node[ylab] at (0, 2.5) {Joint inference of time-varying age-specific contact rates and epidemic state};
\node[ylab] at (0, 1.5) {Time-varying age-specific contact rates modelled with probability distribution};
\node[ylab] at (0, 0.5) {Fixed (pre-epidemic) age-contact structure with inferred scaling for interventions};

\tikzset{citept/.style={blue, font=\tiny, anchor=west, xshift=0.02cm}}
\tikzset{dot/.style={blue, fill=blue, circle, inner sep=1pt}}
\node[dot] at (0.2, 2.8) {}; \node[citept] at (0.2, 2.8) {\textcite{pooley2022estimation}};
\node[dot] at (0.2, 2.6) {}; \node[citept] at (0.2, 2.6) {\textcite{munday2023evaluating}};
\node[dot] at (0.2, 2.4) {}; \node[citept] at (0.2, 2.4) {\textcite{shaman2012forecasting}};
\node[dot] at (2.05, 2.8) {}; \node[citept] at (2.05, 2.8) {\textcite{tran2021sars}};
\node[dot] at (0.2, 1.8) {}; \node[citept] at (0.2, 1.8) {\textcite{manna2023generalized}};
\node[dot] at (0.2, 1.6) {}; \node[citept] at (0.2, 1.6) {\textcite{kucharski2014contribution}};
\node[dot] at (0.2, 1.4) {}; \node[citept] at (0.2, 1.4) {\textcite{britton2025improving}};
\node[dot] at (0.2, 1.2) {}; \node[citept] at (0.2, 1.2) {\textcite{hilton2024beta}};
\node[dot] at (0.2, 0.8) {}; 
\node[citept] at (0.2, 0.8) {\textcite{prem2017projecting}};
\node[dot] at (0.2, 0.6) {}; 
\node[citept] at (0.2, 0.6) {\textcite{prem2021projecting}};
\node[dot] at (2.05, 0.8) {}; \node[citept] at (2.05, 0.8) {\textcite{sonabend2021non}};
\node[dot] at (2.05, 0.6) {}; \node[citept] at (2.05, 0.6) {\textcite{barnard2022modelling}};
\node[dot] at (2.05, 0.4) {}; \node[citept] at (2.05, 0.4) {\textcite{keeling2022comparison}};
\node[dot] at (2.05, 0.2) {}; \node[citept] at (2.05, 0.2) {\textcite{perez2023epidemiological}};
\node[dot] at (3.9, 2.8) {}; \node[citept] at (3.9, 2.8) {\textcite{rosa2025transmission}};
\node[dot] at (3.9, 1.8) {}; \node[citept] at (3.9, 1.8) {\textcite{cocucci2022inference}};
\node[dot] at (3.9, 0.8) {}; \node[citept] at (3.9, 0.8) {\textcite{jewell2023bayesian}};
\node[dot] at (5.7, 1.8) {}; \node[citept] at (5.7, 1.8) {\textcite{cocucci2022inference}};
\node[dot] at (5.7, 1.6) {}; \node[citept] at (5.7, 1.6) {\textcite{smieszek2011reconstructing}};
\node[dot] at (5.7, 1.4) {}; \node[citept] at (5.7, 1.4) {\textcite{viboud2018rapidd}};
\node[dot] at (5.7, 0.8) {}; \node[citept] at (5.7, 0.8) {\textcite{meyer2017incorporating}};
\end{tikzpicture}
\vspace{6pt}
\caption{\textbf{Mapping of different efforts to account for heterogeneity and uncertainty in connectivity matrices in the existing literature.} Reviewed papers are listed according to model stratification (x-axis) and inference methods (y-axis), to identify research gaps. There are notably very few studies that perform joint inference of the connectivity matrix and epidemic state, and none that do so with fine-grained spatial stratification and age-stratification.}
\label{figure:mapping_existing_studies}
\end{figure*}

Beyond data availability, the incorporation of uncertainties and heterogeneities into connectivity matrices remains challenging. Current standard practice typically relies on a fixed connectivity matrix as a model input \cite{sonabend2021non, keeling2022comparison, barnard2022modelling, perez2023epidemiological}, an approach that accounts for neither sampling uncertainty nor individual-level variability in contact behaviour. More recent advances, including probabilistic approaches \cite{hilton2024beta}, joint inference of contact structure and epidemic state \cite{pooley2022estimation, tran2021sars, munday2023evaluating}, data assimilation \cite{shaman2012forecasting, cocucci2022inference, rosa2025transmission}, and ensemble approaches \cite{viboud2018rapidd}, have begun to address these limitations (Figure \ref{figure:mapping_existing_studies}), though issues of structural and practical parameter identifiability for epidemic models remain an unexplored topic \cite{wieland2021structural, browning2020identifiability, simpson2021profile, minas2019parameter}.

Given the growing reliance on connectivity matrices in infectious disease modelling, a review of the data used to construct them and the methods developed to account for uncertainty and heterogeneity is both timely and necessary. This study reviews the existing literature on these topics, identifies key methodological challenges, and highlights directions for future research. Section \ref{sec:data used} examines data types, quality and availability used to construct connectivity matrices; Section \ref{sec:method} reviews methods for incorporating uncertainty and heterogeneity; Section \ref{sec:in practice} discusses constructing connectivity matrices in practice; and Section \ref{sec:discussion} summarises key findings and future directions.

\section{Data Types Used to Construct Connectivity Matrices, and Their Quality and Availability} \label{sec:data used}
Typical approaches to constructing connectivity matrices require empirical data on interactions between individuals. Broadly, two types of empirical data are used: 1) data that directly records the contacts between individuals and 2) data that acts as a proxy for the contacts. In scenarios with limited or no contact data, information on demographics (age, gender, occupation, etc.), geography (location, setting type), or behaviour (compliance with intervention policies) is required to construct the connectivity matrix. 
 
Another classification of empirical data is whether it is collected during an epidemic (i.e., real-time data) or under baseline conditions (i.e., historical data). During an epidemic, interventions and increased awareness of the disease alter social contact patterns compared to normal baseline conditions. Constructing connectivity matrices during an epidemic from baseline data is challenging because of changes in contact patterns, demographics, and behaviour. Even using data from a historical epidemic to the current one is not straightforward, as it poses the same challenge of adapting connectivity matrices to shifts in contact patterns, demographics, and behaviour. 

\subsection{Empirical Data on Interactions}
Generally, empirical contact data is collected using two approaches. First, diary-based surveys, in which a sample of individuals from the population is recruited to record their interactions with others over a period, along with demographic, geographic, and behavioural information. The diary-based survey can be in the form of online questionnaires or a physical diary that participants need to fill throughout the day, or at the end of the day, or the next day, in which case it is known as a retrospective survey. Diary-based survey data is also known as self-reported, egocentric data. Second, digital contact data, in which technologies such as Bluetooth or Radio-Frequency Identification (RFID) tags are used to electronically record the interaction event, with participants recording demographic and behavioural data related to the contact. 

Both data collection approaches have distinct advantages and limitations that we will discuss later in this section. A common limitation across both approaches is the sampling or recruitment of individuals for the survey and the choice of the data collection period \cite{pooley2022estimation,klepac2020contacts,prem2017projecting,kiti2023changing}. Obtaining a representative sample of the population is critical; a biased sample can introduce systematic errors in constructing connectivity matrices, thereby biasing model predictions and data inference. Depending on the scope of the study, sample representativeness is based on either one factor (demographic, geographic, etc.) or a combination of factors. Another aspect is the temporal representativeness of the data, including day of the week, time of day, month, season, etc. The data collection time period influences the frequency of contacts between individuals; if not accounted for, it can lead to over- or under-reporting. Hence, it is essential to account for the sampling assumptions while constructing connectivity matrices. Finally, another limitation is that data collected describes the behaviour of individuals when they are healthy, despite the purpose generally being informing transmission, which occurs (largely, though not exclusively) when individuals are ill \cite{van2013impact}.  

In addition to challenges and limitations related to data quality, data availability remains an important issue in constructing connectivity matrices \cite{dunbar2024transmission}. Foremost, the lack of data from low-income countries leads to reliance on contact patterns inferred from high-income settings, which do not represent the underlying behaviour and can bias modelling outcomes \cite{hoang2019systematic}. Another one is the lack of data on socioeconomic status \cite{barreras2024exciting} and ethnicity \cite{goodfellow2025post}, which can lead to systematic misrepresentation of contact patterns and transmission dynamics in underrepresented populations \cite{oliver2020mobile}. The proxy datasets, such as mobility data, are usually collected by private organisations, which limit accessibility due to privacy and data protection concerns. Also, the cost of procuring, storing, and analysing the data limits the data accessibility. The frequent unavailability of region-specific survey data in the public domain further restricts independent validation across studies \cite{oliver2020mobile, barreras2024exciting}. 

Although all empirical data sources are subject to constraints in availability and quality, the assumptions required to translate them into connectivity matrices differ fundamentally. Survey data rely on self-reporting and sampling assumptions, digital data depend on technological accuracy and adoption, and proxy data require indirect inference across contexts. Understanding these data-specific limitations and quality issues is essential for assessing the usability of each source in constructing the connectivity matrices. We therefore review the challenges associated with survey, digital, and proxy data in turn.

\subsubsection{Survey Data}

Social contact surveys are used to collect data on the number of contacts participants have with individuals in different population groups. The POLYMOD study was the first large-scale effort to systematically collect social contact data across eight European countries \cite{mossong2008social}. Since then, many national and subnational studies with similar designs have been conducted in both pre-pandemic and pandemic settings; a systematic review on these surveys is provided in \cite{hoang2019systematic}.
To capture changes in contact behaviour during the pandemic, the CoMix study was conducted (following the same format as the POLYMOD study) across multiple European countries \cite{verelst2021socrates,jarvis2020quantifying, wong2023social, gimma2022comix, coletti2020comix, backer2023dynamics} incorporated additional behavioural and contextual variables related to public health interventions and mobility. More recently, the ReCoNnect project aimed to characterise post-pandemic social contact patterns in the United Kingdom during 2024–2025 \cite{goodfellow2025post}. Several of these survey datasets can be accessed openly through the \href{https://www.socialcontactdata.org}{Social Contact Data Hub}. Additional datasets are archived in associated Zenodo collection \href{https://zenodo.org/communities/social_contact_data}{Social contact data}. 
                 
Typically, the population is subdivided into different age groups, but the groups may represent a range of different characteristics that affect risk, such as geographic location, sex, socioeconomic status, and contact level. Then a contact needs to be defined in a way that is deemed epidemiologically meaningful for the infection at hand: a common one for directly transmissible airborne infections, like POLYMOD and derived studies, is a two-way conversation in the physical presence of another person, but slight variations or multiple definitions (non-physical, physical, etc.) are considered \cite{mousa2021social}. These surveys then provide information on the total number of contacts $t_{ij}$ each individual in a particular group $i$ had with individuals in group $j$ over a defined time period (typically the previous 24 hours). Dividing this by the number of individuals in the survey in group $i$, $s_i$, gives the mean number of contacts each individual in group $i$ had with individuals in group $j$, $m_{ij}=t_{ij}/s_i$. This is the raw connectivity matrix $M = (m_{ij})$ from the survey.

The numbers of individuals in different groups in the survey do not necessarily reflect their relative proportions in the population as a whole, so we need to account for this when estimating the connectivity matrix $C = (c_{ij})$, which gives the mean number of contacts $c_{ij}$ each individual in group $i$ has with individuals in group $j$ in the whole population. To do this, we generally appeal to the principle of reciprocity (assuming contacts are defined in a symmetric way, like a two-way conversation), so that when individual A in group $i$ contacts individual B in group $j$, individual B also contacts individual A, so the total number of contacts group $i$ has with group $j$ must be equal to the total number of contacts group $j$ has with group $i$. If the numbers of individuals in group $i$ and group $j$ are $n_i$ and $n_j$, respectively, this corresponds to:
$$c_{ij}n_i=c_{ji}n_j.$$

We can satisfy this condition on $c_{ij}$ by supposing that the total number of contacts of individuals of group $i$ with individuals of group $j$ in the population is the average of the total number of contacts of group $i$ with group $j$ in the population based on what was reported, $m_{ij} n_i$, and the total number of contacts of group $j$ with group $i$ had the contacts instead been reported by individuals of group $j$, $m_{ji} n_j$:
$$\frac{(m_{ij}n_i+m_{ji}n_j)}{2}.$$
Dividing this by $n_i$ then gives $c_{ij}$:
$$c_{ij}=\frac{1}{2}\frac{1}{n_i}(m_{ij}n_i+m_{ji}n_j).$$

\paragraph{Limitations of survey data}
Despite widespread use of diary-based survey data, their suitability for constructing reliable connectivity matrices is subject to important limitations. Connectivity matrices derived from self-reported diaries rely heavily on assumptions about accurate recall, consistent interpretation of what constitutes a contact, and representative sampling across population sub-groups. In practice, incomplete or inaccurate reporting of contacts and demographic attributes introduces measurement error that propagates directly into estimated contact rates. Moreover, violations of reciprocity \cite{hamilton2022failure}, where reported contacts between two population groups are not symmetric, are frequently observed, necessitating post hoc adjustments that can substantially alter the matrix structure and introduce additional uncertainty \cite{deolia2025data, prem2017projecting, weerasuriya2022updating, meyer2017incorporating, klepac2020contacts}.

The retrospective design of most diary-based surveys further complicates contact-matrix estimation, as recall bias disproportionately affects short, casual, or repetitive contacts that may nevertheless be epidemiologically relevant. In addition, the reporting burden increases with the number of contacts participants have. As a result, participants may choose not to complete the survey, report contacts for a different day with fewer interactions, or omit some contacts from their responses \cite{nixon2022mixed}. This behaviour can lead to an under-representation of contacts in traditional social contact surveys, compared to digital collection, where the burden on the participant does not increase with the number of contacts.  

The lack of a standardised survey is observed across studies \cite{willem2020socrates,hoang2022exploring,angeli2024acquires}, including differences in recruitment strategies, sampling frames, and recorded contact attributes, which limit the comparability of the resulting matrices and undermine efforts to combine or transfer contact patterns across settings. As a result, connectivity matrices constructed from different diary-based surveys may reflect methodological artefacts as much as genuine differences in social mixing patterns.

Additional structural limitations restrict the interpretability of diary-derived connectivity matrices. Most surveys capture contacts over a single day or a short observation window, providing only a static snapshot of contact behaviour and limiting their ability to represent temporal variability in mixing patterns \cite{klepac2020contacts, leung2023social, prem2017projecting,hoang2019systematic,rodiah2023age}. Children \cite{klepac2020contacts} and older adults are frequently under-represented, and proximity-based interactions in enclosed spaces are poorly captured, further biasing matrix estimates. Collectively, these limitations highlight that connectivity matrices derived from diary-based surveys are contingent on strong underlying assumptions and that the uncertainties introduced at the data and construction stages propagate directly into epidemic model outputs, requiring careful disclosure when interpreting model results.

\subsubsection{Contact Data: Digital Collection}
Digital contact data collection offers an alternative to self-reported surveys by using electronic devices or mobile applications to record proximity events between individuals \cite{sociopatterns_datasets}. A prominent example is the BBC Pandemic study \cite{klepac2018contagion}, conducted between September 2017 and December 2018, which collected contact and mobility data from a substantially larger sample than earlier diary-based studies such as POLYMOD. Data were gathered via a mobile application that recorded participants’ demographic characteristics, including age, household size, gender, and occupation, as well as location information at hourly intervals. Users also logged contact events, reporting the contact's age, the type of interaction, the location, and whether the contact was new or repeated.

Despite their high temporal resolution and reduced reliance on self-reporting, digital contact datasets present important data quality challenges for constructing connectivity matrices. Proximity detection technologies such as Bluetooth can register signals even when individuals are separated by physical barriers or are not engaged in face-to-face interaction, introducing noise into the inferred connectivity matrix. As a result, digitally recorded proximity events do not necessarily correspond to epidemiologically meaningful contacts, and additional assumptions are required to translate raw signals into contact rates suitable for matrix construction.

Several structural limitations further constrain the use of digital contact data for contact-matrix estimation. Matrix construction depends critically on the accuracy of devices in estimating distance and contact duration, both of which vary across hardware and environmental conditions. The deployment and maintenance of dedicated tagging devices can be costly, limiting sample size and population coverage. When mobile phone GPS data are used to supplement proximity information, privacy and ethical concerns often restrict data access and spatial resolution. Moreover, many digital contact studies are conducted in geographically limited settings, such as specific institutions or regions \cite{machens2013infectious,sociopatterns_datasets}, which constrains the generalisability of the resulting connectivity matrices beyond the study context. Taken together, these limitations indicate that connectivity matrices derived from digital contact data are highly sensitive to technological assumptions and study design choices.

\subsection{Proxy Data and Combining Multiple Data Sources} 
Proxy data are frequently used to infer connectivity matrices when direct measurement of interpersonal interactions is unavailable. These sources include census and other aggregated demographic data, mobility and transport datasets, disease surveillance, and, in some cases, proximity sensor records. For example, the Demographic and Health Survey (DHS) \cite{prem2017projecting} provides information on household structure, population age composition, labour force participation, school enrolment, pupil-to-teacher ratios, and other socio-demographic indicators across multiple low- and middle-income countries. The World Bank \cite{prem2017projecting} provides similar aggregated demographic indicators for countries not covered by DHS, though household-level data are often missing. Mobility data from platforms such as Google and Apple \cite{nouvellet2021reduction,prestige2025estimating} capture real-time or historical population movement across various transport modes and locations, including workplaces, residential areas, retail and recreational spaces, and parks. In addition, disease-specific surveillance systems, such as Google Flu Trends \cite{shaman2012forecasting} or national influenza monitoring programs \cite{bingchen2022contact}, quantify population-level trends through search behaviour or reported cases. Some studies also incorporate proximity sensor data, including Bluetooth or radio-frequency identification (RFID) devices and mobile apps \cite{cencetti2021digital}, to indirectly capture interaction patterns.

Despite their utility, proxy datasets present several limitations for constructing reliable connectivity matrices. Many census and demographic datasets are updated infrequently \cite{sullivan2024quantifying}, leading to missing or outdated information that can bias estimates of population structure. Mobility data often suffer from signal sparsity or incomplete coverage, and disease-specific surveillance data may not generalise to other pathogens \cite{bingchen2022contact}, limiting their transferability to new epidemic contexts. Proximity-based datasets, when used as proxies, may only partially capture relevant contact behaviours and are often restricted to small spatial or temporal windows. Collectively, these factors introduce uncertainty into connectivity matrices derived from proxy sources, making them sensitive to the assumptions required to convert population-level or indirect indicators into age- or location-specific contact rates.

Given these limitations, connectivity matrices constructed from proxy data should be interpreted cautiously. Explicitly acknowledging the assumptions underlying matrix construction and quantifying associated uncertainties are essential to avoid misleading conclusions when using these matrices as inputs for epidemic modelling.

\subsubsection{Mobility and Spatial Data}

During the COVID-19 pandemic, social contact patterns changed rapidly in response to perceived infection risk and the implementation of non-pharmaceutical interventions such as physical distancing measures and travel restrictions. Although several contact surveys were conducted in some countries during the pandemic \cite{coletti2020comix, jarvis2020quantifying, backer2023dynamics, bridgen2022social, zhang2020changes, brankston2021quantifying, feehan2021quantifying, latsuzbaia2020evolving}, such surveys are time- and resource-intensive and must be repeated frequently to capture evolving behaviour. Mobility data therefore emerged as an important complementary source of information, providing high-frequency indicators of population movement and activity, which can be linked indirectly to transmission dynamics \cite{oliver2020mobile, nouvellet2021reduction, prestige2025estimating, tomori2021individual,shubin2024influence}. 

Mobility data are commonly obtained from three broad sources \cite{hu2021human}: public transportation systems (e.g., rail and air passenger records), anonymised mobile network operator data derived from cellular activity, and application-based GPS data from location-enabled services (e.g., navigation apps). For privacy and accessibility reasons, such data are typically released in aggregated form. A widely used example is Google’s Community Mobility Reports \cite{googlemobility}, which quantify changes in visits to, and time spent at, several categories of locations such as workplaces, retail and recreation venues, grocery and pharmacy stores, transit stations, residential areas, and parks.

Unlike contact surveys, mobility data do not observe interpersonal encounters directly. Their contribution is indirect, requiring further processing to translate movement patterns into changes in contact intensity. One common strategy is to derive mobility-based scaling factors and apply them to baseline (pre-pandemic) synthetic connectivity matrices. \textcite{prestige2025estimating}, for example, compared four such approaches. Two approaches mapped Google mobility indicators directly into multiplicative scaling factors, assuming contacts to be proportional to mobility (or to mobility squared), where mobility is expressed as a deviation from a baseline (e.g., 0.5 indicating half the baseline activity). The remaining approaches fitted linear or quadratic regression models linking mobility to average contact rates measured in pandemic-era surveys, defining the scaling factor as the ratio of the fitted contact rate to its baseline value. Similarly, \textcite{di2024mobility} adjusted workplace connectivity matrices for France by using Google workplace mobility as a proxy for changes in workplace attendance. 

Mobility data also provides a natural basis for characterising spatial connectivity. For example, mobile phone data can be aggregated into origin–destination matrices describing daily flows between spatial units \cite{oliver2020mobile}. Beyond mobile phone data, commuting flows have long been used to parametrise metapopulation connectivity. \textcite{jewell2023bayesian} constructed a residence-workplace connectivity matrix across UK districts using commuting volumes and embedded it into the force of infection, such that transmission risk in each area depended partly on infectious prevalence elsewhere weighted by commuting intensity.
Beyond within-country commuting matrices, international passenger movement data can be used to construct inter-country mobility matrices.
\textcite{shubin2024influence} constructed inter-country mobility matrices for the neighbouring Nordic
countries of Denmark, Finland, Norway and Sweden (including travel to and from the rest of the world), distinguishing short-term commuting from longer-term travel. These flows were then embedded in a multi-country transmission model to examine cross-border transmission pathways. Similarly, \textcite{adekunle2020delaying} incorporated international flight volume data from the Official Aviation Guide of the Airways into a stochastic meta-population SEIR model to quantify the impact of travel bans on COVID-19 importation into Australia. 

\subsubsection{Pathogen Genetic Data}
Increasingly, pathogen genome sequencing is becoming standard as a public health response, and this allows new methods for the estimation of connectivity matrices. The sequences are not generated in isolation and are usually associated by what evolutionary epidemiologists simply call metadata. This metadata includes factors like sampling location, host species and host age. These constitute the discrete groups we have discussed thus far, and connectivity can be assessed between the different discrete levels of these metadata variables. The interpretation of connectivity in this approach is slightly different, however. Instead of rates of contact between the different groups, connectivity here instead measures the rate of movement of lineages of the pathogen between the different groups. To use the notation presented here, instead of $c_{ij}$ representing the average number of contacts that each individual in group $i$ has with individuals in group $j$ per unit time, for genetic data it instead represents the average number of movements from group $i$ to group $j$ per unit lineage-time. Lineage-time is closely related to person-time, such that one unit of real time corresponds to $n$ units of lineage-time, where $n$ is the number of distinct lineages present in the pathogen population over that period.

There are two closely-related standard approaches to estimating connectivity matrices from pathogen genetic data, with both contained in the phylodynamic framework, where a latent tree describing the relationship between the pathogen isolates is inferred. Due to the assumed existence of this latent tree structure, phylogeny-based approaches can be applicable to poorly surveilled systems, where none of the movement events were directly observed: conditional on the tree structure and the observed group levels at the tips, movement events can be inferred. 

In order to understand the difference between the two standard methods for estimating the connectivity matrices, a basic understanding of the structure of Bayesian phylodynamic models is required. We construct a Bayesian phylodynamic as a decomposition of a joint process generating trees and discrete traits. Firstly, we define a probabilistic model on the discrete traits observed at the tips, $X$, using a series of continuous time Markov chains with parameters, $\phi$, conditional on a given phylogenetic tree, $T$. These discrete traits constitute the columns of the aligned genetic sequences of the pathogen, and any other discrete traits that are being modelled as changing on the tree. We call the probability of the discrete data under this model our phylogenetic likelihood, $P(X|T,\phi)$. We then define a (potentially marked) tree-valued stochastic process with parameters $\theta$, such that we can evaluate the density of a tree under this process. These stochastic processes are generally stochastic population models, such as pure birth process \cite{yule1925purebirth}, which implicitly define trees when births and deaths are tracked. We call the probability density of our latent tree under this process our genealogical likelihood, $P(T|\theta)$. Finally, we define priors over $\phi$ and $\theta$, $P(\phi,\theta)$. Then our posterior, $P(T,\phi,\theta|X)$, is proportional to the product of these components. The two main approaches for estimating the between-group connectivity matrix differ by whether the contact rates are estimated using a model based on the phylogenetic likelihood or as part of the tree-valued stochastic process defining the genealogical likelihood.

The approach that models the group as a discrete trait evolving along the tree under a continuous time Markov chain in a Bayesian framework, commonly called discrete phylogeography, was developed by \textcite{lemey2009bayesian} building on the work of \textcite{felsenstein1981ML} and \textcite{pagel1994CTMC}. As the name implies, it was developed for spatial movements, but, as was noted at the time, this is simply a special application of model-based ancestral state reconstruction, and the approach is applicable for any discrete trait that could plausibly transition between states. This approach has been fruitfully applied to a wide variety of infectious disease systems, from human viruses (e.g. \cite{duplessis2021phylogeography}) to livestock pathogens (e.g. \cite{jacquot2017bluetongue}) and wildlife diseases (e.g. \cite{luzzago2016chamois}). 

The second approach models the group as part of a marked tree-valued stochastic process. The class of available processes is, theoretically, very broad \cite{king2025genealogy}. When the discrete traits are modelled in this way, they can allow the demographics of the pathogen to differ across levels of the discrete trait. For example if location were our discrete trait, the tree could be generated by coupled location-specific SIR models, which would capture in the structure of the tree the non-linearities introduced by differential susceptible depletion in different locations, and spatial movement of pathogen lineages. However, in practice, the stochastic processes used for discrete trait connectivity inference are limited to the small subclass where closed forms or good approximations are available for the density of the tree under the process, such as the structured coalescent \cite{demaio2015structuredcoalescent} and multi-type birth-death \cite{kuhnert2016mtbd}. Extending the class of practically usable processes is an active research area.

\subsubsection{Combined Data Sources}

With an increasing plethora of data sources available, combining data sources into a single connectivity matrix can be a way of using the benefits of each data source to collectively better inform the connectivity matrix generated. Previous research has been done on when and how to best to combine disparate data sources more broadly in infectious disease modelling \cite[see][for more details and further references]{deangelis2015multipledata,swallowestimationchallenge}.

For connectivity matrices specifically, (a subset of) the different types of data summarised already in this section may be combined to inform different processes. \textcite{feehan2021quantifying} developed a method that integrated survey data, demographic statistics, and mobility records to construct refined and dynamic connectivity matrices. These were used to study contact rate changes across waves of Covid-19 in the US and across demographic heterogeneity.

\textcite{sarratea2026estimation} used age-structured incidence data from early in an outbreak to estimate age-structured connectivity matrix entries from an SIR model. The authors directly relate the contact parameters to various epidemiological quantities that can be estimated through different sources and select between different model structures based on varying assumptions, such as sparsity, age-heterogeneity and case-reporting rate. They also suggest how additional data, such as age-stratified serology, may be able to better inform which scenarios are realistic. The authors apply their method to data from Greater Buenos Aires (Argentina) in 2020, using an individual-based SIR model to test the appropriateness and limitations of the approach when strong restrictions are in place.

\subsection{Demographic and Epidemiological Data}

Demographic data are primarily used to define the baseline connectivity structure between population subgroups, through quantities such as age distribution, population density, and spatial distribution. This information can be used to complement or project survey data to different population structures. For example, the age distribution of a population enables the parametrisation of age-specific matrices, capturing assortativity in mixing across age groups \cite{prem2017projecting, prem2021projecting, arregui2018projecting}. Similarly, population density can be used to inform contact patterns across different urban and rural settings \cite{read2014social}, whereas epidemiological data, such as case counts, hospitalisations, and mortality, are typically used to estimate transmission parameters operating on this contact structure and refining parameters such as effective contact rates, susceptibility and infectiousness through model calibration \cite{shadboltdatachallenge,swallowestimationchallenge}.

\section{Accounting for Uncertainty in Connectivity Matrices} \label{sec:method}

\subsection{Sources of Uncertainty} \label{sec:method-uncertainty}

\begin{figure*}[htbp]
    \centering
    \includegraphics[width = 1 \textwidth]{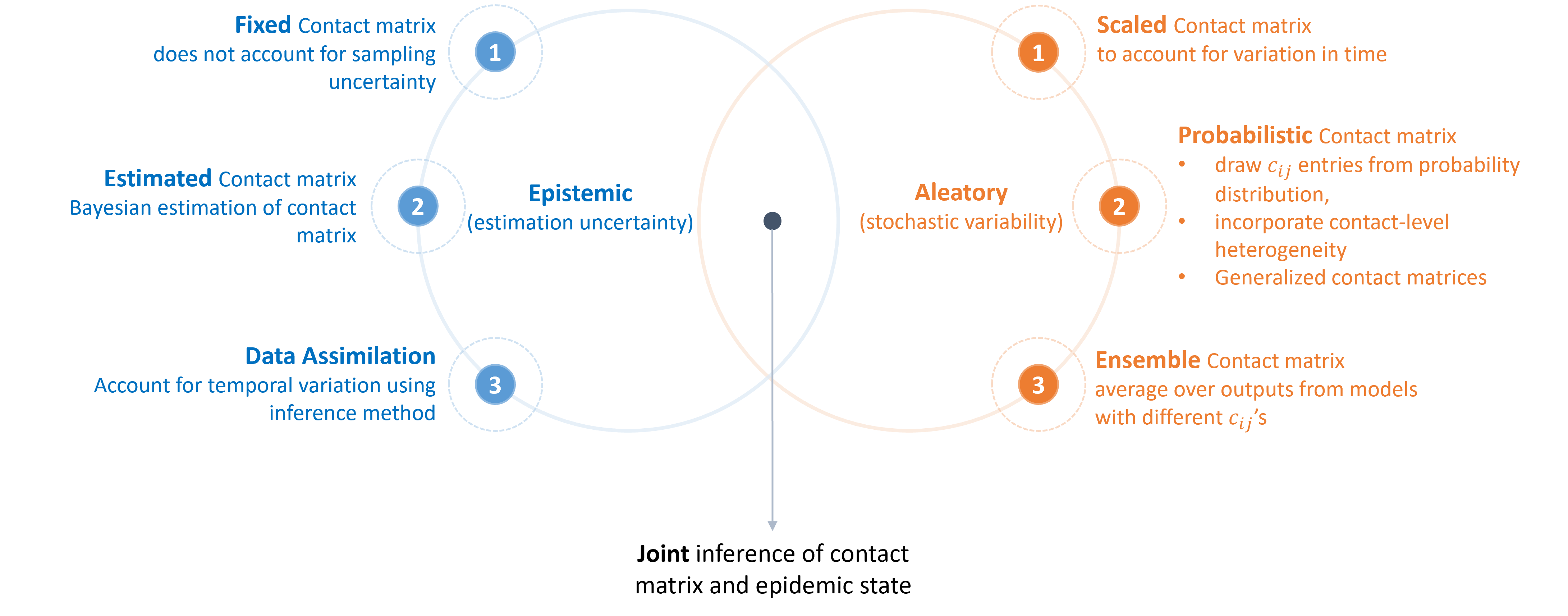}
    \caption{\textbf{Modelling and inference approaches to incorporate uncertainty in connectivity matrices}. Many existing studies use fixed connectivity matrices, despite multiple sources of uncertainty that should be accounted for. These uncertainties can be broadly classified into two types: epistemic uncertainty (reducible uncertainty in the connectivity matrix arising from lack of knowledge, incomplete data, or limitations in model structure) and aleatory uncertainty (inherent irreducible uncertainty in the connectivity matrix due to randomness in the system). A wide range of methods has been proposed to address them; we organize these in terms of the modelling and inference approach used and specify which types of uncertainty each approach tries to address.}
    \label{figure:inference approaches}
\end{figure*}

Connectivity matrices are subject to two types of uncertainty (Figure \ref{figure:inference approaches}): aleatory uncertainty reflecting stochastic variation in human contact behaviour, and epistemic uncertainty arising from incomplete knowledge and imperfect data \cite{bevan2022uncertainty}.

Aleatory uncertainty comes from the inherent stochasticity and heterogeneity in human contact behaviour. This unavoidable variability manifests in several ways. First, individual-level heterogeneity: different people have fundamentally different contact patterns, some individuals consistently maintain many contacts while others have few, creating overdispersion in contact distributions \cite{volz2011effects}. Second, short-term temporal fluctuations: day-to-day randomness in whether specific contacts occur, even for the same individual. Third, longer-term temporal variability: weekly cycles (e.g.~weekdays vs weekends), seasonal effects (school terms, holidays, weather), and contact restrictions during outbreaks. Fourth, as a consequence, variation in contact duration and intensity: contacts between individuals vary in length and physical proximity rather than all being equal, as is generally assumed. 

In contrast, epistemic uncertainty comes from our limited knowledge and measurement capabilities when constructing a connectivity matrix. This uncertainty arises from multiple sources. Sampling uncertainty: finite sample sizes in contact surveys mean that the observed contact patterns may not perfectly represent the true population-level mixing \cite{mossong2008social}. Projection uncertainty: many countries lack direct observations so require projection of contact patterns, and modellers need to make assumptions about how demographic factors, such as school, work, household structure, relate to contact patterns \cite{prem2017projecting}. Structural uncertainty, about modelling choices, and assumptions about how setting-specific contacts combine \cite{hamilton2024examining}. Network observation also brings uncertainty, since some of the network structure of contacts is partially or completely unobserved \cite{wang2024accounting}. 

Aleatory variability cannot be eliminated but can be characterized and incorporated into models, for example, through probability distributions that capture dispersion in contact rates across individuals. Epistemic uncertainty can be reduced by using inference-driven approaches, to estimate unknown contact patterns from available data while quantifying remaining uncertainty.

\subsection{Using a Fixed/Scaled Connectivity Matrix} \label{sec:method-fixed}

A standard approach taken to using connectivity matrices in transmission models when modelling epidemics is to treat the connectivity matrix itself as fixed and use point estimates based on survey data for the mean numbers of contacts between groups. Typically, a whole non-epidemic connectivity matrix $\tilde{c}_{ij}$ is scaled by a constant factor $\alpha$ to get the epidemic contact matrix $c_{ij}$, i.e.~$c_{ij} = \alpha\, \tilde{c}_{ij}$, to account for any difference in the non-epidemic and epidemic contact rates. This does not account for uncertainty in the estimation of the connectivity matrix stemming from only a sample of the population having been surveyed, or changes in contact patterns between subgroups (e.g.~age groups) in response to the epidemic or interventions (e.g.~school closures, movement restrictions), but does allow changes in the overall contact rate during an epidemic to be incorporated, including those over time (by making the scaling factor time-dependent, i.e.~$\alpha_t \tilde{c}_{ij}$). It is the approach that was adopted by most modelling groups during the COVID-19 pandemic, not least because of a lack of real-time data with which to estimate pandemic connectivity matrices for most countries. Examples of this include: modelling of the impact of non-pharmaceutical interventions and vaccination in the UK \cite{sonabend2021non}, modelling of the roadmap out of lockdown in the UK in 2021 \cite{keeling2022comparison}, modelling the closure of red-light areea in Indai \cite{pandey2020effectextendedclosureredlight}, modelling the effects of interventions in the Philippines \cite{caldwell2021understanding}, modelling the epidemic dynamics under Omicron subvariants \cite{barnard2022modelling}, the transmissibility and severity of different SARS-CoV-2 variants \cite{perez2023epidemiological}, and real-time spatial modelling of transmission dynamics \cite{jewell2023bayesian}. \textcite{jewell2023bayesian}, for instance, used commuting data from the 2011 UK census to parameterise flow between local authority districts and a Bayesian framework to infer parameters and unobserved events (infections) with a temporal random effect that accounted for variation in overall contact rates over time and a spatial random effect that absorbed any errors in the interaction between regions based on the census data.

\subsection{Accounting for Sampling Uncertainty} \label{sec:method-sampling}

Sampling uncertainty in the mean numbers of contacts between groups can be estimated by fitting a statistical model to the contact survey data or bootstrapping the survey data. A prime example of this is the work of \cite{prem2017projecting}, who used a Bayesian hierarchical model to estimate connectivity matrices across 152 countries. This approach is particularly useful for dealing with sparse or missing data, as it allows the pooling of information across similar countries or regions while accounting for differences. The Bayesian hierarchical framework effectively incorporates uncertainty and variability in contact patterns by assigning a probability distribution to the contact rates, which varies at different hierarchical levels. At the top level, global data on contact patterns is used to inform the model, while at lower levels, country-specific data is integrated to refine the estimates for each country. In this model, the variability between countries is handled by introducing prior distributions on the parameters, and the final connectivity matrix is obtained by sampling from these distributions. This approach allows the model to capture between-country variability, like differences in social structures, producing more adaptable and accurate connectivity matrices. However, this approach can become computationally intensive and difficult to interpret, especially when integrating data from multiple sources.

In principle, once sampling uncertainty has been estimated, it can be accounted for by running multiple simulations of the epidemic model with different samples of the connectivity matrix drawn from its posterior distribution/bootstrap sampling distribution to generate a distribution of possible outcomes rather than a single outcome. However, this has rarely been done in practice \cite{van2022learning,vizi2025age,korir2025eigenvector,machens2013infectious}.

\subsection{Accounting for Heterogeneity in Contact Between Individuals} \label{sec:method-heterogeneity}

One approach to capturing contact variability is through the use of probability distributions. In a probabilistic connectivity matrix, each entry of the matrix, rather than being a fixed number, is drawn from a probability distribution that reflects the variability in the number of contacts between two groups across individuals. For example, $c_{ij}\sim\operatorname{Poiss}(\mu_{ij})$ or $c_{ij}\sim\operatorname{NegBin}(\mu_{ij},\sigma_{ij}^2)$, where $\mu_{ij}$ and $\sigma_{ij}$ are the mean and, in the negative binomial case, the standard deviation of the number of contacts an individual in group $i$ has with an individual in group $j$. This variability can then be captured in more detailed or individual-level models of transmission by sampling numbers of contacts from the probability distributions. \textcite{hilton2024beta}, for example, proposed a beta-Poisson model that separates variability in contact rates and infection probabilities. This model assumes a Poisson process governs the contact events, while the transmission probability per contact follows a beta distribution. This dual-distribution model provides greater flexibility in modelling interventions that affect both contact behaviour and transmission likelihood, offering more realism in dynamic transmission scenarios. However, the above approaches rely heavily on empirical data to estimate parameters, which may not always be available or reliable. Moreover, the assumptions made by these models may oversimplify the actual complexity of social interactions.


Another approach to incorporate variability in contact rates between individuals is to split individuals in each group according to their contact levels, i.e.~add another level of stratification to the model beyond age/spatial region. Each entry of the connectivity matrix $c_{ik,jl}$ then represents the mean number of contacts between individuals of the respective contact levels $k$ and $l$ within the original two groups $i$ and $j$. \textcite{kucharski2014contribution} used this approach to derive a final size model for influenza outbreaks stratified by age and contact-level. They fitted the model to data from the 2009 Hong Kong A/H1N1 pandemic to examine how much age and contact-level heterogeneity each shape transmission dynamics. \textcite{britton2025improving} used a similar idea, splitting age groups into individuals with high contact levels (“socially active”) and those with low contact levels (“socially inactive”) to look at the impact of different assumptions about assortativity of mixing by age and contact level on $R_0$ and final epidemic size. 

\textcite{manna2023generalized} took this approach a step further by making the indices in the connectivity matrix represent intersections of subgroups in the population to allow for greater population stratification (e.g.~by age, sex, location, ethnicity, etc.). They made the indices vectors $\boldsymbol{i},\boldsymbol{j}$, i.e.~$C=c_{\boldsymbol{i}\boldsymbol{j}}$, such that $\boldsymbol{i}$ could say represent individuals of a certain age, income, ethnicity and education status (rather than just one of these factors), and called the resulting matrices \textit{generalised contact matrices}. They showed that failing to account for such heterogeneity when modelling epidemics could lead to inaccurate representation of transmission dynamics and epidemic outcomes, and to underestimation of $R_0$. However, it could be argued that over-stratification could break the asymptotics required to tend to the dominant eigenvector and therefore needs some further consideration.

\subsection{Joint Inference of the Connectivity Matrix and Epidemic State \label{jointinf}}

An alternative approach treats the elements of the connectivity matrix, or scaling factors applied to them, as random variables on which prior distributions based on existing knowledge or contact survey data can be placed. Bayesian inference can then be used to update these distributions as new data becomes available, such as real-time contact survey data, mobility data or case counts, resulting in a posterior distribution that propagates remaining uncertainty through the transmission model. 

\textcite{pooley2022estimation} used an Approximate Bayesian Computation model-based proposal (ABC-MBP) inference scheme to jointly infer the time-varying reproduction number and the age-stratified connectivity matrix from public COVID-19 data for England. In their paper, the authors use a scaled version of the pre-pandemic connectivity matrix, constructed from the BBC Pandemic survey \cite{klepac2018contagion}. The approach allows the model to increase or reduce the relative rate of contacts for different age groups, scaling rows and columns of the matrix to allow for in-pandemic connectivity to be estimated. Time-dependence during the pandemic is estimated as an overall scaling of the connectivity matrix to represent changes in contact behaviour linked to risk perception and response to non-pharmaceutical interventions. 

Similarly, \textcite{tran2021sars} use detailed age-specific testing and hospitalisation data from all 13 regions of metropolitan France to inform parameters determining age-specific changes in connectivity matrices across regions during the post-lockdown rebound of summer 2020. The authors use a Bayesian MCMC approach to estimate transmission model parameters of a deterministic age-stratified compartmental model, along with scaling of contacts to understand who was driving transmission and to test different age-targeted interventions. A sensitivity analysis is also conducted to vary assumptions about the relative infectivity and susceptibility of the different age groups. 

\textcite{munday2023evaluating} used a renewal equation approach and Hamiltonian Monte Carlo (HMC) to jointly fit the elements of the connectivity matrix to contact data from the CoMix survey \cite{jarvis2020quantifying}, infection incidence estimated from the UK national infection and antibody prevalence survey \cite{ONS2022}, and reported COVID-19 cases. By placing Gamma priors on the elements of the connectivity matrix, they were able to infer a time-varying connectivity matrix. Their work found that including real-time contact data improved forecasting predictions at relatively short intervals compared to using pre-pandemic contact data or inferring the connectivity matrix purely from the epidemiological data, but this improvement varied across the phase of the epidemic.

\subsection{Data Assimilation Approach} \label{sec:method-assimilation}

An alternative to a static matrix being updated with data in joint inference is to use data assimilation methods, such as the Kalman filter or particle filters, to adjust the connectivity matrix sequentially as new mobility data or disease incidence data become available. When timely connectivity data is accessible this updating can be performed in real-time. Data Assimilation methods revise both the epidemic state and the contact structure at each time step by combining the model forecast with incoming observations, making it possible to track changes in contact patterns in real time as an outbreak evolves.

\textcite{shaman2012forecasting} illustrates the use of data assimilation techniques, by applying the Ensemble Kalman Filter (EnKF), in real-time epidemic forecasting of seasonal influenza, which can be used to continuously update the connectivity matrix. The authors develop an inference procedure that incorporates retrospective ensemble forecasts generated on a weekly basis following assimilation of web-based estimates of influenza-like illness. While this application adjusted a baseline transmission rate rather than a structured matrix, it established the methodological foundation for applying sequential data assimilation to epidemic models and demonstrated that transmission dynamics can be estimated and continuously revised in real time as new observations arrive.

More recent work has extended this framework to explicitly update structured connectivity matrices as part of the data assimilation step. \textcite{cocucci2022inference} coupled the EnKF to an agent-based epidemic model in which the contact network was constructed from demographic and mobility data describing household, school, and workplace interactions between individuals. Building on similar principles, \textcite{rosa2025transmission} treated the elements of an age-structured transmission matrix as time-dependent variables within the EnKF, estimating their values sequentially from reported COVID-19 age-stratified case data. Rather than scaling a fixed pre-pandemic connectivity matrix, the filter was allowed to learn the changing interaction rates between age groups directly from the data, with the matrix entries updated at each assimilation step as new observations arrived. The approach was validated through experiments using synthetic data before application to real-world observations, confirming that the EnKF could recover time-varying matrix entries under controlled conditions. The authors note, that the simultaneous estimation of all matrix elements poses identifiability challenges, and that reliable recovery requires appropriate constraints on the parameterisation of the matrix.

The feasibility of data assimilation for connectivity matrices depends on the availability of connectivity/mobility data that can serve as real-time proxies for changing contact patterns. Several studies have demonstrated the value of such data, providing methodological context for the data assimilation approaches described above. \textcite{tizzoni2014use} explores the use of agent-based models for scenario analysis in epidemic modelling, illustrating how streaming mobility and incidence data can be integrated online to continuously update estimates of effective transmission across population groups. Similarly, \textcite{smieszek2011reconstructing} uses an individual-based model in which contact rates are parameterized with probability distributions derived from empirical contact data, demonstrating how variability in contact patterns can be modelled using statistical distributions that are compatible with particle-filter update. Finally, \textcite{meyer2017incorporating} integrate age-structured social connectivity matrices into spatio-temporal endemic–epidemic models for surveillance counts, providing a principled way to connect contact data to transmission between groups; such frameworks can be used to update contact-driven parameters as new surveillance data become available or to provide structured priors for data assimilation.

\subsection{Ensemble Approach} \label{sec:method-ensemble}

It is increasingly common for a suite of models to be available with different paramaterisations, structures and/or data assimilations. Each of these models could have a different connectivity matrix based on different assumptions or data sources. The ensemble average or distribution of outcomes can provide an estimate that incorporates the uncertainty in the connectivity matrices.

\textcite{viboud2018rapidd} develop an ensemble forecast on the RAPIDD Ebola Forecasting Challenge, a large collaborative exercise designed to evaluate how well different infectious disease models can forecast epidemics under controlled conditions. Four different scenarios were tested, combining predictions from different models and using different performance metrics. The challenge used synthetic Ebola outbreak data, inspired by the 2014–2015 West African Ebola epidemic, allowing systematic comparison of models without the confounding issues of real-world data gaps. The authors demonstrate how combining different models (and by extension, different connectivity matrices) can better account for uncertainties. 

\textcite{riley2007large} discusses the importance of using multiple models to account for different aspects of disease transmission, including varying connectivity matrices. In particular, household dynamics often require a small scale, simple model that can be extended to other spatial scales and distances. The author highlights that no single model can capture all aspects of an epidemic, and using an ensemble enhances the robustness of predictions, such as those used during the 2001 foot-and-mouth disease outbreak in the United Kingdom.

\section{Constructing Connectivity Matrices in Practice} \label{sec:in practice}

In the previous sections, we discussed the primary data sources used to estimate connectivity matrices, together with data quality and availability challenges. We also reviewed state-of-the-art methods, tools and approaches to enable estimation of connectivity matrices. In this section, we discuss the practical complexities of using available data to estimate connectivity matrices and the consequent influence on modelling outcomes.

\subsection{Using Connectivity Matrices in Real-time} 
We first consider scenarios in which connectivity matrices are informed by real-time data, in which case the contact rate varies over time. Real-time contact data enables the incorporation of real-time behavioural changes associated with interventions, risk perception, susceptibility, immunity, policy compliance, etc. \cite{munday2023evaluating, prestige2025estimating, feehan2021quantifying}. However, during the early phase of an epidemic, real-time contact data are often sparse, delayed, or geographically limited, reducing their identifiability. As a result, real-time informed connectivity matrices may provide only limited additional information beyond baseline assumptions during the initial growth phase of an outbreak. 

In contrast, historical contact data collected before the epidemic are frequently used to inform connectivity matrices. For epidemic modelling, the connectivity matrix is typically adapted to the current setting (region, disease, etc.) by global or setting-specific scaling parameters \cite{sullivan2024quantifying, van2022augmenting, mistry2021inferring}. While this approach supports qualitative inference, mismatches can arise due to changes in population demographics, occupational patterns, mobility behaviour, technology-mediated interactions, and behavioural responses to disease risk. Empirical comparisons \cite{jarvis2020quantifying} between pre-pandemic and during-pandemic contact patterns indicate that there are significant changes in social contact patterns, particularly for age-specific and setting-specific contacts.
 
Some demographic changes can be partially accommodated by using proxy data, such as census-derived population structure, school records, workplace records, or mobility indicators \cite{prestige2025estimating}, through weighted aggregation or stratified scaling \cite{mistry2021inferring}. However, these adjustments cannot fully capture intervention-induced behavioural shifts or changes in contact heterogeneity. Studies \cite{jarvis2020quantifying} have shown that using historical connectivity matrices can bias modelling outcomes, including age-specific attack rates, thereby affecting inference on epidemic size, the reproduction number \cite{feehan2021quantifying,munday2023evaluating}, and the targeting of interventions.

In the third scenario, when no contact data are available, the modelling approach treats the connectivity matrix as a latent variable inferred from transmission parameters, using epidemic outcome data such as case counts, hospitalisations, or serological surveys. Here, connectivity matrices are constrained by strong prior assumptions, leading to weakly identifiable matrices, i.e., multiple structurally distinct matrices may generate similar epidemic trajectories. The model outputs are highly sensitive to priors, limiting the quantitative interpretability of the predictions.

Following the separate examination of the challenges associated with real-time, historical, and no contact data, we note that the study by \textcite{munday2023evaluating} discussed in section 3\ref{jointinf} evaluates the use of these three data sources. The authors compare the results of a forecasting model trained on real-time data (CoMix survey), historical data (POLYMOD), and no contact data for COVID-19 in the UK. Results show that when fitting models to case data influenced by testing capacity or reporting practices, estimating the connectivity matrix from either real-time or historical data does not improve forecasting performance over the no-contact-data model. In contrast, when fitting to infection incidence data inferred from prevalence data collected using a standardised method in the UK COVID-19 Infection Survey \cite{ONS2022}, the model informed by a real-time connectivity matrix consistently outperforms models with historical or inferred connectivity matrices for short- and long-term forecasts.

The above findings show that the temporality and granularity of the connectivity matrix do not uniformly translate into improved model performance. The utility of real-time, historical, or inferred connectivity matrices depends on the epidemic phase, data consistency, and the degree of identifiability from available observations. In some settings, simple assumptions may provide more robust inference than structurally misaligned or weakly informed connectivity matrices.

\subsection{Identifiability in Connectivity Matrix Estimation} \label{sec:identifiability}
Identifiability is an important general issue for any kind of parametric modelling and is certainly pertinent for epidemic models featuring connectivity matrices. On the one hand, such models typically have sufficiently many parameters to make it unclear as to whether or not all parameters could be estimated even in an idealised setting of complete observation of an epidemic. In addition, real-life levels of observation raise similar questions of what can or cannot be estimated in practice. These two scenarios are concerned with two kinds of identifiability as we now briefly explain.

First, a model parameter (or set of parameters) is structurally identifiable if distinct values of the parameter give rise to distinct realisations of the model. For a deterministic model a realisation is its single trajectory, while for a stochastic model a realisation here refers to the probability distribution of all possible trajectories. 
Conversely, a lack of structural identifiability means that different values of the parameter can give rise to identical model realisations. A trivial example would be a (single-type) model in which an infection rate from subgroup $j$ to subgroup $i$ was parameterised as $\beta_{ij} = p\, c_{ij}$, so any choices of $p$ and $c_{ij}$ giving the same $\beta_{ij}$ value would give identical model realisations. Parameter structural identifiability can be equivalently defined as the property in which perfect noiseless observation of the model enables precise estimation of the parameter in question. 

The second kind of identifiability is practical identifiability. A model parameter is practically identifiable if it can be accurately estimated given available data. Practical identifiability is therefore concerned not only with the model itself, but also the kind of observational data to hand. Both forms of identifiability are also closely related to parameter sensitivity, here referring to the extent that changes to one or more parameters cause changes in model realisations.

There is an extensive literature on identifiability, much of which is concerned with structural identifiability in deterministic models, particularly with applications in biology \cite[see e.g.][and references therein]{wieland2021structural}. Methods have also been developed to assess practical identifiability and parameter sensitivity for stochastic models (typically stochastic differential equation models, e.g., \cite{browning2020identifiability, simpson2021profile, minas2019parameter}). However, identifiability for epidemic models has received far less focus in the literature, although attention has been drawn to some of the inherent subtleties and challenges of the interplay between such models and data \cite[see e.g.][]{cunniffe2024identifiability,sauer2022identifiability,kiss2023parameter}. In particular, identifiability for epidemic models featuring connectivity matrices remains a largely unexplored topic that is of both theoretical and practical interest.

\section{Discussion}\label{sec:discussion}

In this review, we have studied the common structures and data used to construct connectivity matrices for informing the spread of infectious diseases in humans. A variety of data types are available that can be used to inform the connectivity between heterogeneous subpopulations. These include demographic and epidemiological data, survey data, mobility and spatial data and genetic data. Each of these data sources can provide information on the interaction between individuals in the population or movements between genetic lineages in the case of phylogenetic data. Increasingly, there is recognition that combining these different sources can improve coverage and account for different components of the connectivity matrices. However, this combination is not without additional challenges \cite{deangelis2015multipledata, swallowestimationchallenge}.

We have also highlighted the possible approaches to combining inference on disease transmission with varying degrees of heterogeneity, uncertainty or measurement error in connectivity matrices. Metapopulation models and other similar patch model structures that allow for heterogeneity within populations often require some form of structure to be enforced on the way that transmission events occur through movement or other dynamic contact processes in order to ensure the information content of the partially-observed data is sufficient for inference. Whilst these processes are important to understand when studying or predicting disease dynamics, they are often over-parameterised and subject to imperfect data, leading to issues of either structural or practical identifiability. These non-identifiabilities are often complex and state-dependent and practical tools to enable careful consideration of their implications are still lacking.

In order to account for either measurement error in or un-measured dynamics in connectivity matrices, it is important to account for some stochasticity in the processes they describe. Methods that account for both aleatory and epistemic uncertainty have been developed and the nature of the model structures may need to depend on the type and quality of data source(s) used to inform them. Various families of approaches have been developed to assist with both the inherent stochasticity in the system as well as the uncertainties in the data and statistical estimation processes. 

Firstly, methods based on scaling data-informed connectivity matrices using data from commuting or other movement surveillance surveys. These methods attempt to allow for changes relative to a baseline survey, which may have been collected under different settings. Similar approaches account for sampling uncertainty or heterogeneity by adding hierarchical data and/or structures to the connectivity matrices.

A second set of approaches treat the matrix or components thereof as random variables with assumed probability distributions. Bayesian statistical techniques are particularly designed to tackle this sort of approach, with methods using filtering and projection equations to account for differing levels of uncertainty and propagate this through the transition model.

Joint inference over both the connectivity matrix and epidemic state process can also be tackled in this way, but can provide challenges when available data are not fully informative on the quantities of interest, leading to practical non-identifiability. Dealing with this problem typically involves assuming some degree of structure on the matrix or providing informative prior distributions, as well as forcing sparsity or temporal smoothness to reduce degrees of freedom in the inverse problem.

So far little research has been done to formally generate ensembles of connectivity matrices, with inferences and predictions being done more commonly across models rather than across connectivity matrices, therefore essentially marginalising out the matrix.

Current standard practice is to use a fixed connectivity matrix based on non-epidemic survey data and scale the whole matrix according to differences in contact rates during the epidemic and over time, and use this as a fixed input in the epidemic model when fitting it to data. This accounts for neither sampling uncertainty nor variability in contact levels between individuals. It is also possible to use a mean matrix, but change the matrix elements based on knowledge of implications of interventions, e.g. when schools are closed, or based on CoMix over time. We instead recommend that the survey data, as a sample of contact rates between individuals in the different groups in the population, is treated as observed data, and used alongside epidemic data to jointly infer the connectivity matrix and the epidemic model state and parameters, to properly account for sampling uncertainty. We also recommend the use of more highly stratified models, e.g.~including stratification by contact level, to account for contact heterogeneity between population subgroups and the differences in epidemic dynamics this can cause. All these must be carefully considered within the identifiability constraints of the model structure and available data.


\section{Acknowledgements}
The programme was supported by EPSRC grants EP/R014604/1 and EP/V521929/1. Additional funding was provided by the Heilbronn Institute Additional Funding Programme for Mathematical Sciences, a £300 million investment over a five-year period from 2020/21 to 2024/25. This new funding is being delivered by EPSRC. GM was supported by UK research and innovation grant BB/W007711/1 and the Scottish Government's Rural and Environment Science and Analytical Services Division (RESAS). DJP was supported by the National Institute for Health and Care Research (NIHR) Cambridge Biomedical Research Centre NIHR203312 grant. EN is affiliated to the NIHR Health Protection Research Unit in Emerging and Zoonotic Infections (NIHR HPRU-EZI) (NIHR207393) at the University of Liverpool in partnership with the UK Health Security Agency (UKHSA), in collaboration with Liverpool School of Tropical Medicine, London School of Hygiene and Tropical Medicine and The University of Oxford. The views expressed are those of the authors and not necessarily those of the NIHR, UKHSA or the Department of Health and Social Care.

The authors would like to thank the Isaac Newton Institute for Mathematical Sciences, Cambridge, for support and hospitality during the programme `Modelling and Inference for Pandemic Preparedness' and its follow-on meeting where work on this paper was initially undertaken.


\printbibliography
\end{document}